\documentclass[journal]{IEEEtran}

\usepackage[hidelinks]{hyperref}
\usepackage{epsfig,amsmath,graphicx,amssymb,xcolor,color,dsfont,caption,subcaption}

\newcommand{\x}{\boldsymbol{x}}

\usepackage{booktabs}
\usepackage{multirow}
\usepackage{tabularx}
\newcolumntype{Y}{>{\centering\arraybackslash}X}
\usepackage{lipsum}
\usepackage[boldmath]{numprint}
\usepackage[ruled,linesnumbered]{algorithm2e}
\ifCLASSINFOpdf
\else
\fi
\begin{document}

\title{Representation Learning With Hidden Unit Clustering For Low Resource Speech Applications}

\author{Varun Krishna,~\IEEEmembership{Student Member,~IEEE,} Tarun Sai,~\IEEEmembership{Student Member,~IEEE,} and \\ Sriram Ganapathy,~\IEEEmembership{Senior Member,~IEEE,}
\thanks{V. Krishna, S. Tarun and S. Ganapathy are with the Learning and Extraction of Acoustic Patterns (LEAP) lab, Department of Electrical Engineering, Indian Institute of Science, Bangalore, India, 560012.
 e-mail: \{varunkrishna, tarunsai, sriramg\}@iisc.ac.in}}
\markboth{Journal of \LaTeX\ Class Files,~Vol.~XX, No.~X, January 2022}%
{Shell \MakeLowercase{\textit{et al.}}: A Sample Article Using IEEEtran.cls for IEEE Journals}


\maketitle

\begin{abstract}
The representation learning of speech, without textual resources, is an area of significant interest for many low resource speech applications. In this paper, we describe an approach to self-supervised representation learning from raw audio
using a hidden unit clustering (HUC) framework.
The input to the model consists of audio samples  that are windowed and processed with  1-D convolutional layers. The learned ``time-frequency'' representations from the convolutional neural network (CNN) module are further processed with long short term memory (LSTM) layers which generate a contextual vector representation for every windowed segment. The HUC framework, allowing the categorization of the representations into a small number of phoneme-like units, is used to train the model for learning {semantically rich} speech representations.  The targets consist of phoneme-like pseudo labels for each audio segment and these are generated with an iterative k-means algorithm.  We explore techniques that improve the speaker invariance of the learned representations and illustrate the effectiveness of  the proposed approach on two settings, i) completely unsupervised speech applications on the sub-tasks described as part of the ZeroSpeech 2021 challenge and ii) semi-supervised automatic speech recognition (ASR) applications on the TIMIT dataset and on the  GramVaani challenge Hindi dataset. In these experiments, we achieve state-of-art results for  various Zerospeech tasks. Further, on the ASR experiments, the HUC representations are shown to improve significantly over other established benchmarks based on {wav2vec,  HuBERT and Best-RQ}.
\end{abstract}


\section{Introduction}
\IEEEPARstart{S}{poken} language processing using raw speech, without any textual resources,  has received wide spread interest recently~\cite{lakhotia2021generative} and are useful for various applications like zero-resource spoken language modeling, low-resource speech recognition and speech synthesis. The key challenge in these tasks is the automatic discovery of sub-word units of speech~\cite{park2007unsupervised} that are speaker invariant and consistent.

The learning of unsupervised audio representations~\cite{dunbar2021zero} has been largely investigated with  self-supervised loss functions involving contrastive learning ~\cite{liu2021self}, vector quantization~\cite{baevski2020vqwav2vec}, clustering~\cite{maekaku21_interspeech} or autoregressive predictive coding~\cite{Chung2019AnUA}.  For unsupervised learning of audio representations, the use of restricted Boltzmann machines \cite{sailor2016filterbank,agrawal2017unsupervised},  variational learning \cite{agrawal2019unsupervised} and generative adversarial networks \cite{agrawal2018comparison} have been explored to learn acoustic/modulation filters. Recently, a family of models, termed as \textit{wav2vec}~\cite{schneider19_interspeech,  baevski2019vq,baevski2020wav2vec}, were investigated for self-supervised representation learning applications like speech recognition and synthesis. The initial  approach~\cite{schneider19_interspeech} used the contrastive predictive coding~\cite{oord2019representation} with  $1$-D convolution layers. The subsequent vector-quantization based approach, \textit{wav2vec-vq} \cite{baevski2019vq}, was inspired by the  quantization module proposed by Oord et. al.~\cite{oord2018neural}. The output of the quantization module is employed in training  bidirectional encoder representations from transformer (BERT)   language models \cite{DBLP:conf/naacl/DevlinCLT19}. The recent approach, \textit{wav2vec 2.0}~\cite{baevski2020wav2vec},   further added a learnable code-book based model.

The ZeroSpeech series of challenges  \cite{versteegh15_interspeech, dunbar2017zero, dunbar20_interspeech, dunbar2021zero} considered an open call for participation in speech representation learning.  The challenges bench-marked various methods for self-supervised learning  based on word similarity and spoken language modeling tasks.  The zero shot metrics used in the ZeroSpeech 2021 challenge \cite{dunbar2021zero} considered acoustic   and linguistic measures.
Several approaches based on Gaussian mixture modeling~\cite{heck2017feature}, HMMs     \cite{ansari2017unsupervised}, and deep learning methods \cite{niekerk20b_interspeech,ansari2017deep} have been explored for the ZeroSpeech tasks.  In the recent years, self-supervised learning methods has also been investigated for ZeroSpeech applications. These efforts have investigated various model architectures like convolutional networks ~\cite{chorowski2019unsupervised}, recurrent networks~\cite{ravanelli2020multi},  transformer models~\cite{liu2021tera}, and conformer models~\cite{maekaku21_interspeech}.
 For spoken language modeling tasks, most of prior works  use a separate clustering model on the embeddings. {Further, most of the prior works generate representations that are general purpose and do not attempt any disentanglement of the factors to enhance the semantic richness of the representations.}

In this paper, we explore a framework of hidden unit clustering (HUC), where the deep representations learned by a neural network are optimized using a categorical loss. This paper extends our previous work on acoustic unit discovery \cite{varunicassp2022}.
The work is inspired by prior work on deep clustering by Caron et. al.~\cite{caron2018deep} for self-supervised learning from the raw audio data. In our work, the HUC model is shown to enable the learning of discriminative representations that are  less impacted by speaker variations.
The experimental validation of the proposed approach is performed on two settings. The first setting is  the ZeroSpeech 2021 challenge task.   The  phoneme recognition experiments are performed on the TIMIT dataset while the large vocabulary speech recognition experiments are performed on the semi-supervised track of  the GramVaani Hindi ASR challenge\footnote{\url{https://sites.google.com/view/gramvaaniasrchallenge/home}}.

The novel contributions from this current work over the previous approach~\cite{varunicassp2022} are,
\begin{itemize}

\item We propose a self-supervised learning approach using a combination of contrastive loss and HUC loss that generates speaker invariant representations.
{The semantic richness in the representations is achieved using an utterance level mean normalization performed  on the context vector representations.}

\item We develop a data sampling technique  to generate the pseudo-phoneme cluster labels. The  labels learned using the sampling approach is shown to be effective in generating categorical representations.

\item We propose a self-supervised dimensionality reduction method on the representations, driven by the pseudo-labels. This is based on gradient boosted decision tree modeling \cite{Chen:2016:XST:2939672.2939785}.

\item  {A set of semantic task evaluations illustrate the performance gains using the proposed framework. Using the data from the ZeroSpeech 2021 challenge, we show that the proposed HUC based self-supervision achieves state-of-art results on $7$ out of the $8$ sub-tasks. } We also show that the representations from the proposed model perform well in phoneme recognition and low-resource ASR experiments.
\item {A detailed analysis on the representations and the evaluations on non-semantic tasks shows that the proposed HUC framework achieves the semantic richness   at the cost of being a general purpose speech representation for non-semantic speech tasks. }
\end{itemize}


\section{Background}\label{sec:background}
\subsection{Related prior work}
\subsubsection{\textbf{Contrastive predictive coding}}\label{sec:cpc}
The contrastive predictive  coding~\cite{oord2019representation} network consists of two sets of processing layers, with inputs being the raw audio samples.
Let $x[n]$ denote the audio samples that are windowed into short-term frames (typically of $20$ms duration). The output of the convolutional block contains  $f$ dimensional filterbank representations for each frame $t$. These outputs, sampled with a stride of $10$ ms, are denoted as $\boldsymbol{z}_t$,  are then fed to the LSTM layer. The LSTM layer uses a context window of $v$ previous frames,  $(\boldsymbol{z}_t,\boldsymbol{z}_{t-1},...,\boldsymbol{z}_{t-v})$. The output of the LSTM layer is referred to as the context vector $\boldsymbol{c}_t$. The contrastive loss is computed based on the prediction of the $K$ future filterbank representations, $\{\boldsymbol{z}_{t+k}\}_{1 \leq k \leq K}$ by minimizing the following loss function~\cite{oord2018neural},
\begin{equation}
    \mathcal{L}_t = -\frac{1}{K}\sum_{k=1}^{K}log \left(\frac{exp({\boldsymbol z}_{t+k}^T{\boldsymbol W}_k{\boldsymbol c}_t)}{\sum_{\Tilde{{\boldsymbol z}}\in\mathcal{N}}exp(\tilde{{\boldsymbol z}}^T{\boldsymbol W}_k{\boldsymbol c}_t)} \right)\label{eq:cpc_loss}
\end{equation}
Here, $\mathcal{N}$ denotes a subset of negative examples chosen randomly, and $\textbf{W}_k$ forms a set of weights used in the prediction of the $k^{\text{th}}$ future embedding $\boldsymbol{z}_{t+k}$.
{The CPC model used for the experiments reported in this paper employs negative sampling from within the same utterance (within speaker negative sampling).}

\subsubsection{\textbf{wav2vec models}}:
The wav2vec \cite{schneider19_interspeech, baevski2020vqwav2vec, baevski2020wav2vec} architecture emulates the 1-D convolutional layer of the CPC model.

\textit{wav2vec}: Instead of the LSTM layer used in the CPC model, the representations $\boldsymbol{z}_t$ are processed with a second set of convolutional layers with a window of $v$ contextual embeddings, $(\boldsymbol{z}_t,\boldsymbol{z}_{t-1},...,\boldsymbol{z}_{t-v})$. The model outputs the contextual vector representation $\boldsymbol{c}_t$, for each time step $t$. The model learns to distinguish a sample $\boldsymbol{z}_{t+k}$ from distractor samples $\tilde{z}$.

\begin{equation}
    \mathcal{L}_k = - \sum_{t=1}^{T-k} log (\sigma ({\boldsymbol z}_{t+k} ^T h_k({\boldsymbol c}_t)) + \lambda \mathop{{}\mathbb{E}}_{\Tilde{\boldsymbol z}\sim p_{n}} log (\sigma (-\Tilde{\boldsymbol z} ^T h_k ({\boldsymbol c}_t)))
\end{equation}
Here, $\tilde{z}$ is drawn from a proposal distribution $p_n$, $h_k$ defines a linear feed-forward neural network layer and {$\sigma(.)$ denotes sigmoid activation function}. Further, the weight factor $\lambda$ is a hyper-parameter.

\textit{wav2vec-vq}:
In this model \cite{baevski2020vqwav2vec}, the quantized context vector representations are used in the  prediction task. {The model architecture resembles that of wav2vec model, with two convolutional networks $\boldsymbol f: \boldsymbol{\mathcal{X}}\mapsto \boldsymbol{\mathcal{Z}} $ and $\boldsymbol g: \boldsymbol{\mathcal{\hat{Z}}}\mapsto \boldsymbol{\mathcal{C}} $ for feature extraction and aggregation, and an additional vector quantization module $\boldsymbol q: \boldsymbol{\mathcal{Z}}\mapsto \boldsymbol{\mathcal{\hat{Z}}} $}


\textit{wav2vec 2.0}:
{
The wav2vec 2.0~\cite{baevski2020wav2vec} builds over the ideas of wav2vec-vq \cite{baevski2020vqwav2vec}. The wav2vec 2.0 uses a multi-layer convolution feature encoder $\boldsymbol{f} : \boldsymbol{\mathcal{X}} \mapsto \boldsymbol{\mathcal{Z}}$ with the raw audio as input. A portion of the features are masked before the transformer  based aggregator $\boldsymbol{g} : \boldsymbol{\mathcal{Z}} \mapsto \boldsymbol{\mathcal{C}}$. The context vectors corresponding to the unmasked regions are used to predict the       quantized \cite{5432202} representations of the masked regions.}

\subsubsection{\textbf{Deep clustering}}
{
The deep clustering approach, proposed for learning visual representations, combines feature learning  and clustering into a single neural model \cite{caron2018deep}.
Unlike the VQ based approach, the deep clustering does not require the context vectors to be approximated as code-book entries. For speech modality, deep clustering has been explored in the HuBERT framework.  }
\subsubsection{\textbf{HuBERT}} {The
HuBERT \cite{hsu2021hubert} model consists of  convolutional layers  followed by a set of transformer encoder layers. During the pre-training stage, transformer \cite{vaswani2017attention} encoder consumes the masked acoustic \cite{joshi2020spanbert} features $\boldsymbol{U}$ from the convolutional encoder. The output $\boldsymbol{H}$ of the transformer is used to predict the discrete targets $\boldsymbol{z}\in [K]$ ($K$-class categorical variable) through a linear projection layer. The HuBERT adopts an iterative clustering and training process. For the first iteration, the mel frequency features, categorized into $\boldsymbol{K}$ clusters,  are  used as discrete targets. For subsequent iterations, discrete targets are derived by clustering the hidden layer representations.}

\begin{figure}[t!]
    \centering
   \includegraphics[width=0.95\columnwidth]{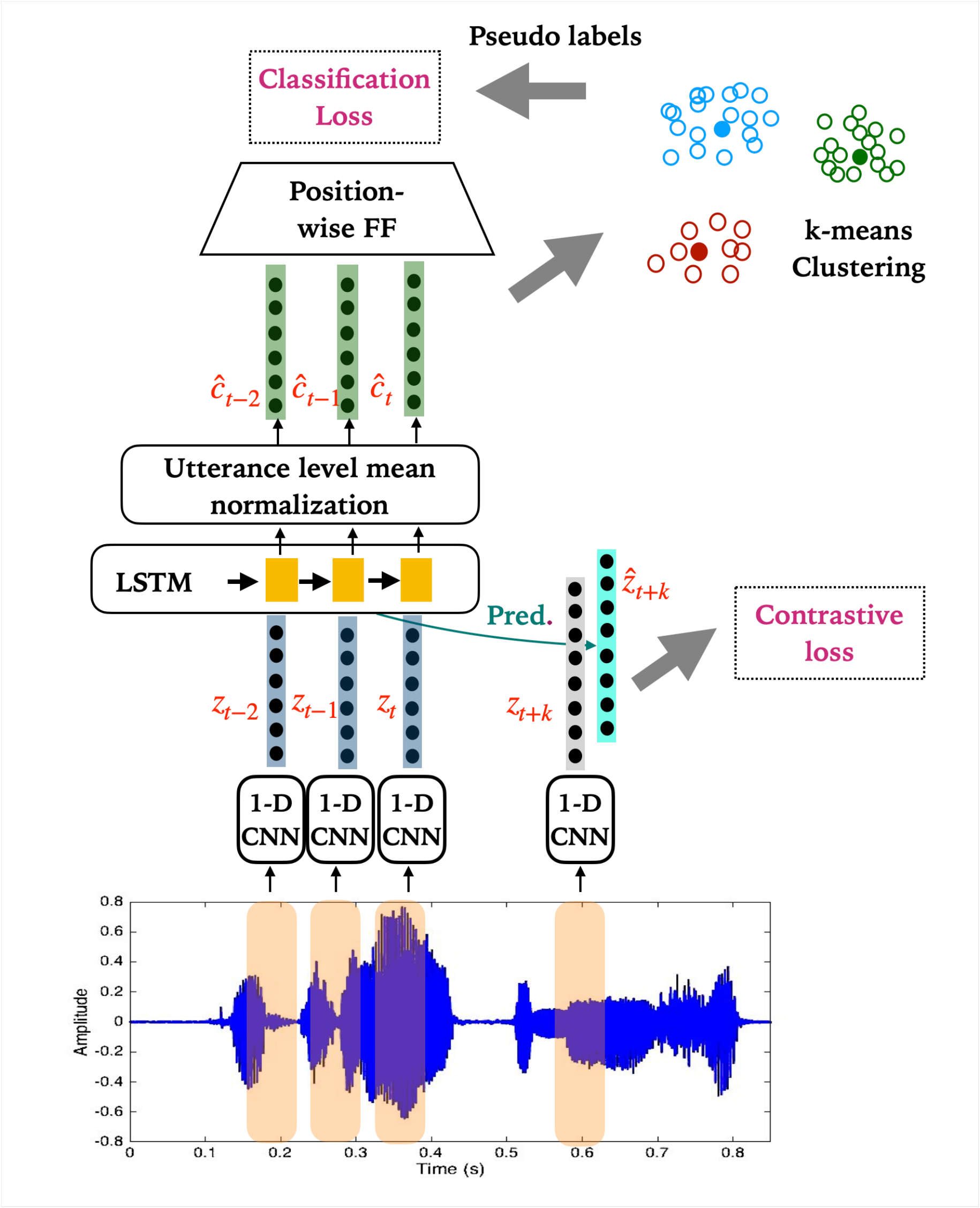}
    \vspace{-0.05in}
    \caption{Proposed approach for learning speaker invariant representations using the HUC framework. }
    \label{fig:block}

\end{figure}
\subsection{Contrast with proposed HUC model}
{Many of the prior works perform self supervised learning (SSL) on the English read speech in Librispeech \cite{Panayotov2015LibrispeechAA} dataset, while illustrating the quality of the representations using speech recognition tasks \cite{baevski2020wav2vec, hsu2021hubert}. In English read speech utterances, the two key streams of information are the semantic (content) variability and the non-semantic (speaker) variability. Most of the prior works do not attempt to explicitly disentangle the semantic and non-semantic information streams as they aim to derive general purpose audio representations. In contrast, the proposed work a) identifies a way to quantify the non-semantic component of a given speech utterance, b) uses the non-semantic component in a diverse sampling of the acoustic units, and, c) factors this component out of the final representations in order to enhance the semantic encoding of speech. We illustrate this contrast experimentally using various downstream tasks that probe the semantic and non-semantic encoding in the speech representations derived from the HUC framework.}

{ Our work, while also using deep clustering,   differs from HuBERT~\cite{hsu2021hubert} in the following ways - i) We do not employ any masking of the audio   regions, ii) A mean normalized representation from the LSTM aggregator is used for clustering, iii) A data sampling framework, based on diversity of pseudo-speaker clusters is used to design the code-books, iv) The model is trained with  a combination of clustering and CPC losses. With various ablation experiments, we empirically highlight the advantages of  these processing steps. }

\section{Proposed work}\label{sec:self-supervised}

The overview of the proposed framework is shown in Fig.~\ref{fig:block}. The initial processing steps consist of a block of 1-D convolution layers, similar to the CPC model~\cite{oord2019representation}.
The convolutional layers generate filter-bank features denoted as  ${\boldsymbol z}_t^i$, where $t$ is the frame index and $i$ is the utterance index.  Subsequently, the LSTM layer generates context vectors  ${\boldsymbol c}_t^i$.
The regularized HUC loss is used to train the  model.

The HUC approach attempts to generate categorical labels using a clustering framework applied on the CPC based context vector representations. For this purpose, the context vectors ${\boldsymbol c}_t^i$ from a pre-trained model are clustered using an unsupervised k-means clustering. The labels from this clustering step are considered to be the psuedo-phoneme labels for training the model shown in Fig.~\ref{fig:block}.

\subsection{Analyzing the mean context vector}
\begin{figure}[t!]
    \centering
   \includegraphics[width=0.9\columnwidth]{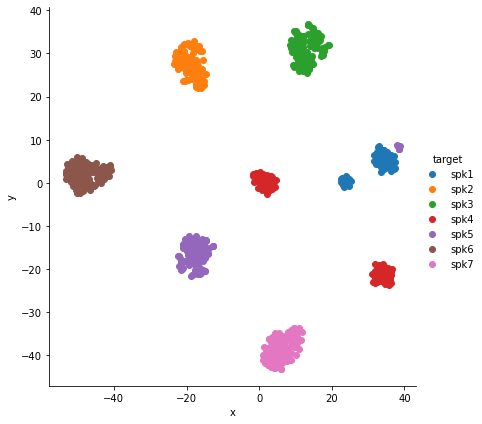}
    \vspace{-0.05in}
    \caption{ The t-SNE plot of the mean context vector,  $\boldsymbol{\Bar{c}^i}$, computed for $50$-$100$ utterances each, from a set of $7$ speakers in the LibriSpeech dataset.}
    \label{fig:tSNE}

\end{figure}

In our analysis, shown in Fig~\ref{fig:tSNE}, we use a pre-trained CPC model {trained} on the Librispeech dataset, {employing within utterance negative sampling of distractors}, to compute the utterance level mean, $\boldsymbol{\Bar{c}}^i$, of the context vector representation. We plot the t-distributed stochastic neighbourhood embedding (t-SNE) \cite{van2008visualizing} of the mean context vectors, for $50-100$ utterances derived from  $7$ speakers chosen randomly from the dataset. As seen in this plot, the utterance level mean of the context vectors primarily captures the speaker information.

\subsection{Data sampling for k-means training}\label{sec:data-sampling}
With a large number of utterances, and with a frame-rate of $100$ Hz ($10$ms windows), the number of  context vectors $\boldsymbol{ {c}}_t^i$ available for clustering is significantly large ($>100M$).
Secondly, the pseudo-phoneme classes (classes for training the HUC model) have to satisfy two goals namely,
\begin{itemize}
    \item \textit{consistency} - should be sufficiently broad to represent different variants of the same underlying sound spoken by different speakers and in different acoustic contexts.
    \item \textit{conciseness} - should be succinct to capture the fine acoustic properties that make it distinctive from other units.
\end{itemize}

The data sampling process attempts to achieve the right trade-off between the consistency and conciseness. The goal of the data sampling process is to create a subset of the training data, that are used in the clustering.
The algorithm is outlined in Alg. $1$.

The key objective is to preserve only the utterances from diverse  ``speakers''  in the corpus. The first step in the data sampling is to perform a k-means clustering of the mean context vectors, $\boldsymbol{\Bar {c}}^i$, derived from all the utterances in the training data. This step generates $M$ pseudo-speaker centroids of the training data. The optimal value of $M$ is determined using knee point algorithm  \cite{satopaa2011finding, sugar2003finding}for k-means. { The knee point algorithm finds the point of inflection in inertia. Specifically, the sum of squared distances of a point from the nearest cluster center is computed for different choices of $M$ (number of clusters). The point of inflection is used as the optimal value for $M$}. Further, using the inter-centroid distance matrix ({the matrix containing the pair-wise distance between the $M$ centroids}), the set of $N$ centroids ($N < M$) are {identified as extreme cluster centroids (the centroids which are maximally distant), as outlined in Alg. 1}.

In the second step, any given utterance $i$ of the training data is mapped to one of the $M$ pseudo-speaker centroids, by generating the mean context vector $\boldsymbol{\Bar {c}}^i$ and finding the closest centroid. The utterances, whose mean context vector  associates to one of the $N$ extreme centroids are alone used in the HUC.  This sampling process thus retains only the maximally diverse utterances for clustering.
The value of $N$ is a hyper-parameter and is obtained through validation experiments.

\begin{algorithm}[t!]
\caption{Training utterance sampling algorithm}
\textbf{Step I}: Extract context vectors from the training data $\big \{ \{\boldsymbol {c}_t^i \}_{t=1}^{T_i}  \big \}_{i=1}^U$, $U$ denotes the total number of training utterances, and $T_i$ denotes the length of the $i$th utterance in terms of the number of frames.  \\
\textbf{Step II}: Compute utterance level means -  $\{\boldsymbol{\Bar {c}}^i\}_{i=1}^U$. \\
\textbf{Step III}: Cluster the utterance level means $\{\boldsymbol{\Bar {c}}^i\}_{i=1}^U$ to generate $M$ centroids using k-means clustering - $\mathcal{C} = \{\{\boldsymbol{\Bar {d}}^k\}_{k=1}^M \}$, where $\boldsymbol{\Bar {d}}$ denotes the pseudo-speaker centroids.  \\
\textbf{Step IV}: Identify a subset of $N$ farthest centroids from $\mathcal{C}$. This set is denoted as $\mathcal{ {C ^\prime}}$.  \\
\textbf{Step V}: For all $i=1..U$, compute the closest centroid using
$   j_*^i = argmin_{j=1..M} ||\boldsymbol{\Bar {c}}^i - \boldsymbol{\Bar {d}}^j ||^2$ \\
\textbf{Step VI}: If $ \boldsymbol{\Bar {d}}^{j_*^i} \in \mathcal{ {C ^\prime}}$, all the mean-removed context vectors $\big \{ \boldsymbol {\hat {c}}_t^i \}_{t=1}^{T_i}$ are selected for training the HUC labels. Else, utterance $i$ is not used in generating the acoustic units.

\end{algorithm}
The sampled set of mean normalized context vectors are input to the k-means clustering. We use the Euclidean distance for the clustering. The clustering process generates the pseudo-phoneme labels for each time frame of the context vector.   The k-means clustering approach is run for $100$ iterations and the final cluster centers are stored. These cluster centers are used to obtain the pseudo-phoneme labels for each frame of  the audio utterance in the training dataset. The model architecture, shown in Fig.~\ref{fig:block}, is trained from random initialization using the frame level pseudo-phoneme labels.

\subsection{Representation pre-processing}
 In order to generate representations that are speaker invariant, we remove the utterance level mean, $\boldsymbol{\Bar{c}}^i$, from the context vectors ${\boldsymbol c}_t^i$ during the HUC model training. The mean normalized context vectors are denoted as $\boldsymbol{\hat {c}}_t^i$.

\subsection{Loss function}

The HUC loss is the cross-entropy loss, where the   logits layer (a linear layer followed by softmax non-linearity) on the mean-normalized context vectors, ${\boldsymbol {\hat c}}_t$, (Fig.~\ref{fig:block}) is used the predict the pseudo-phoneme labels.    We define the  cluster loss as,
\begin{equation}{\label{eq:cross_Entropy}}
    \mathcal{L}_C = - \sum _{t=1}^{T} {\boldsymbol l}_t ^T log (S(f(\boldsymbol{\hat {c}}_t))),
\end{equation}
where $S$ denotes the softmax function, $f$ denotes the {logits layer (a linear layer with soft-max non-linearity)},  and ${\boldsymbol l}_t$ denotes the one-hot encoded label vector (cluster identity) for the context vector ${\boldsymbol c}_t$. The {logits} layer allows the non-linear mapping of the context vectors ${\boldsymbol c}_t$. As the clustering is a non-linear process, the  {logits layer} transformation with soft-max activation enables the model to learn the cluster labels.

The number of clusters $k$ defines the granularity of the categorical space. A small value of $k$ reduces the number of target classes in the clustering network and allows for a faster training of the models. However, a larger value of $k$ allows the model to capture the finer acoustic categories. Our experiments indicate that a value of $k=200$ provides the best performance in terms of the ABX metric \cite{schatz2013evaluating} on the ZeroSpeech $2021$ challenge dataset.

We also explore a regularization of the HUC loss with the CPC loss (defined in Eq.~\ref{eq:cpc_loss}).  The regularized loss is given by,
\begin{equation}\label{eq:reg-loss}
 \mathcal{L}_{reg} =  \mathcal{L}_{\mathcal{C}}   + \lambda \mathcal{L}_{cpc}
\end{equation}
Here, $\lambda$ is the regularization factor.
The clustering loss is a frame-level loss while the CPC loss encourages the prediction of future frames conditioned on the previous time-frames.  Thus, the regularized loss allows the benefits of categorical learning (from the clustering loss) without over-fitting.
In our experiments, we have chosen $\lambda << 1$ to emphasize the importance of the clustering loss in the representation learning framework.




\subsection{Dimensionality reduction}\label{subsec:JM_mean}
For the ABX tasks, we further enrich the categorical nature of the context vector representations, $\boldsymbol{c}_t^i$, using a discriminative dimensionality reduction approach.  This is achieved by retaining only those  dimensions that are the most predictive of the pseudo-phoneme labels. We build a gradient boosted decision tree model  \cite{Chen:2016:XST:2939672.2939785} with the inputs being the context vector representations, $\boldsymbol{c}_t^i$ and the targets being the corresponding pseudo-phoneme labels.
The decision tree model    {\cite{Chen:2016:XST:2939672.2939785}} allows the ranking of the feature dimensions and the top $D$ dimensions are retained.


\section{ZeroSpeech experiments}\label{sec:zero-experiments}
\subsection{Data}
The training data is derived from LibriSpeech~\cite{Panayotov2015LibrispeechAA} and Libri-light dataset~\cite{DBLP:journals/corr/abs-1912-07875}.
The LibriSpeech corpus contains data from audio-books that are part of the LibriVox project, and it consists of $1000$ hours of read speech sampled at $16$kHz. The training data is derived from more than $1000$ speakers, almost equally split across the two genders, with each speaker providing $25-30$ minutes of audio.
The Libri-light is also derived from open-source audio books from the LibriVox project. It contains over $60$K hours of audio. In the ZeroSpeech 2021 challenge  \cite{dunbar2021zero}, the data is provided without any textual or speaker information.

\subsection{HUC model training}
The  CPC-big~\cite{dunbar2021zero} model, trained on the Librispeech $960$h  dataset, generates the initial context vector representations.
The encoder architecture of the CPC-big model contains $5$ layers of  $1$-D convolutional networks with kernel sizes of $10$,$8$,$4$,$4$,$4$ and stride of $5$,$4$,$2$,$2$,$2$, respectively. The autoregressive model is a $4$ layer LSTM network with a hidden dimension  of $512$ units.

The encoder architecture of the HUC model is similar to the CPC-big model. However, $2$ layers of the LSTM network, with a hidden dimension of $256$ units, are used for the autoregressive modeling.  The HUC model is trained with the regularized  loss function, defined in Eq.~\ref{eq:reg-loss}. The number of pseudo-speaker clusters $N$, HUC units $k$, and regularization factor $\lambda$ are hyper-parameters, which are set using the ABX metric on the ZeroSpeech development data.
The pseudo phoneme labels extracted from the full librispeech data (LS-960h) are used to train the HUC model. The models are trained for $200$ epochs with a patience factor of $5$.

For all the other tasks, like the language modeling on the ZeroSpeech dataset and the ASR experiments, the hyper-parameters are not modified.   Following the HUC model training, the pseudo-phoneme sequences, obtained as the frame-level class predictions from the model, extracted on the LS-960h data, are used for training the BERT based language model.

\subsection{Performance metrics}
The metrics defined in the Zerospeech-2021 challenge \cite{dunbar2021zero} are computed on the development and test set. The results reported in the paper are for Track-1 - speech based language modeling task.

\noindent \textbf{Phonetic}~\cite{schatz2013evaluating}: Given a triplet of tri-phone words,  $\textbf{A}$,$\textbf{B}$ and $\textbf{X}$, where $\textbf{A}$ and $\textbf{B}$ differ  in the center phoneme, while $\textbf{A}$ and $\textbf{X}$ are two different utterances of the same word, the ABX metric computes the fraction of instances when $\textbf{A}$ and $\textbf{X}$ are more distant than $\textbf{A}$ and $\textbf{B}$. {
The average angular distance of the representations along the dynamic time warped (DTW) path is used to compute the distance between the utterances of the words.
The ABX score computed when all the tri-phone words are chosen from same speaker is called the ``within'' speaker ABX, while the case when the words $\textbf{A}$  and $\textbf{B}$ are derived from the same speaker, while $\textbf{X}$ is chosen from  a different speaker is called the ``across'' speaker ABX. Experiments are also reported on the ``clean'' and ``other'' splits of the  Librispeech development data \cite{Panayotov2015LibrispeechAA} data.  }   \newline
\textbf{Lexical}~\cite{le-godais-etal-2017-comparing}: The sWUGGY "spot-the-word" measures the ability of the model to identify a legitimate word. A set of word/non-word pairs are selected. The measure is computed as the fraction of the pairs where the likelihood of the legitimate word is higher than non-word. Since it is an accuracy metric, a  higher value is preferred. \newline
\textbf{Syntactic}~\cite{warstadt2020blimp}: The sBLIMP metric measures the ability of model to identify grammatically correct sentences. It is computed as fraction of instances where the  likelihood of grammatically correct sentence is higher than an incorrect one. Thus, a higher value is preferred. \newline
\textbf{Semantic}: The sSIMI metric is used to assess the lexical semantics and is computed as the Spearman’s rank correlation coefficient $\rho$ between the semantic similarity scores between given by the model and the human scores in the dataset.

{The phonetic task (ABX) uses the representations from the self-supervised learning (SSL) models to compute the ABX metric. For the rest of the tasks, the ZeroSpeech $2021$ challenge \cite{dunbar2021zero} proposes a cascading of the self-supervised feature extractor module,  k-means based quantizer and a language model. The k-means quantizer, trained on the representations from the feature extractor, is used to discretize an utterance. These discretized sequences are used to train a BERT-based language model \cite{dunbar2021zero}. During the inference,  the likelihood score of the discretized sequence, given by the language model, is used for the spoken language modeling tasks. }

\begin{table}[t!]
\centering
{
\caption{ABX error (\%) metric on ZeroSpeech development data for different values of regularization factor $\lambda$. HUC units are set to $k=50$ }
\begin{tabular}{@{}cl|cccc@{}}
\toprule
\multicolumn{2}{l|}{\multirow{2}{*}{\textbf{Regularization}}} & \multicolumn{4}{c}{\textbf{ABX}}                                                                                                                                                                                                                                                                                                                          \\ \cmidrule(l){3-6}
\multicolumn{2}{l|}{}                                         & \multicolumn{1}{c|}{\textbf{\begin{tabular}[c]{@{}c@{}}Clean \\ Within\end{tabular}}} & \multicolumn{1}{l|}{\textbf{\begin{tabular}[c]{@{}l@{}}Clean\\ Across\end{tabular}}} & \multicolumn{1}{l|}{\textbf{\begin{tabular}[c]{@{}l@{}}Other\\ Within\end{tabular}}} & \multicolumn{1}{l}{\textbf{\begin{tabular}[c]{@{}l@{}}Other\\ Across\end{tabular}}} \\ \midrule
\multicolumn{2}{c|}{$2\mathrm{e}-1$}                                     & \multicolumn{1}{c|}{7.57}                                                             & \multicolumn{1}{c|}{10.46}                                                           & \multicolumn{1}{c|}{9.47}                                                            & 16.05                                                                               \\
\multicolumn{2}{c|}{$1\mathrm{e}{-1}$}                                      & \multicolumn{1}{c|}{5.83}                                                             & \multicolumn{1}{c|}{8.21}                                                            & \multicolumn{1}{c|}{7.71}                                                            & 13.06                                                                               \\
\multicolumn{2}{c|}{$1\mathrm{e}{-2}$}                                     & \multicolumn{1}{c|}{4.42}                                                             & \multicolumn{1}{c|}{6.34}                                                            & \multicolumn{1}{c|}{6.58}                                                            & 11.73                                                                               \\
\multicolumn{2}{c|}{$1\mathrm{e}{-3}$}                                    & \multicolumn{1}{c|}{3.86}                                                             & \multicolumn{1}{c|}{5.32}                                                            & \multicolumn{1}{c|}{6.14}                                                            & 10.49                                                                               \\
\multicolumn{2}{c|}{$1\mathrm{e}{-4}$}                                   & \multicolumn{1}{c|}{\textbf{3.72}}                                                    & \multicolumn{1}{c|}{\textbf{5.08}}                                                   & \multicolumn{1}{c|}{\textbf{5.94}}                                                   & \textbf{10.26}                                                                      \\
\multicolumn{2}{c|}{$1\mathrm{e}{-5}$}                                  & \multicolumn{1}{c|}{3.80}                                                             & \multicolumn{1}{c|}{5.16}                                                            & \multicolumn{1}{c|}{6.11}                                                            & 10.38                                                                               \\
\multicolumn{2}{c|}{$1\mathrm{e}{-6}$}                                 & \multicolumn{1}{c|}{3.94}                                                             & \multicolumn{1}{c|}{5.21}                                                            & \multicolumn{1}{c|}{6.08}                                                            & 10.33                                                                               \\
\multicolumn{2}{c|}{0}                                        & \multicolumn{1}{c|}{3.82}                                                             & \multicolumn{1}{c|}{5.14}                                                            & \multicolumn{1}{c|}{6.05}                                                            & 10.43                                                                               \\ \bottomrule
\end{tabular}
\label{tab:ABX1}
}
\end{table}

\begin{table}[t!]
\centering

\caption{ABX error (\%) metric on ZeroSpeech development data for different values of the number of clusters $k$. Regularization parameter $\lambda$ is set to $1\mathrm{e}{-10^4}$}
\begin{tabular}{@{}c|cccc@{}}
\toprule
\multirow{2}{*}{\textbf{\# clusters}} & \multicolumn{4}{c}{\textbf{ABX}} \\ \cmidrule(l){2-5}
($k$)
 & \multicolumn{1}{c|}{\textbf{\begin{tabular}[c]{@{}c@{}}Clean \\ Within\end{tabular}}} & \multicolumn{1}{c|}{\textbf{\begin{tabular}[c]{@{}c@{}}Clean \\ Across\end{tabular}}} & \multicolumn{1}{c|}{\textbf{\begin{tabular}[c]{@{}c@{}}Other \\ Within\end{tabular}}} & \textbf{\begin{tabular}[c]{@{}c@{}}Other \\ Across\end{tabular}} \\ \midrule
50 & \multicolumn{1}{c|}{3.72} & \multicolumn{1}{c|}{5.08} & \multicolumn{1}{c|}{5.94} & 10.26 \\
200 & \multicolumn{1}{c|}{\textbf{3.60}} & \multicolumn{1}{c|}{\textbf{4.82}} & \multicolumn{1}{c|}{\textbf{5.71}} & \textbf{9.15} \\
400 & \multicolumn{1}{c|}{4.28} & \multicolumn{1}{c|}{5.63} & \multicolumn{1}{c|}{6.15} & 9.90 \\ \bottomrule
\end{tabular}
\label{tab:ABX2}
\end{table}

\begin{table}[t!]
\centering
\caption{ABX error (\%) on the ZeroSpeech 2021 development set for different values of $N$ (number of farthest pseudo-speaker clusters chosen). The last row indicates results when  $N=30$ closest pseudo-speaker clusters are chosen instead of the farthest. {Experiments use $\lambda=1\mathrm{e}{-4}$ and $k=200$}. Norm. represents mean normalization. {We also compare with other choices of sampling the data.}}
\begin{tabular}{@{}l|cccc@{}}
\toprule
\multirow{2}{*}{\textbf{\# pseudo-speaker }} & \multicolumn{4}{c}{\textbf{ABX}} \\ \cmidrule(l){2-5}
\textbf{clusters} ($N$) & \multicolumn{1}{c|}{\textbf{\begin{tabular}[c]{@{}c@{}}Clean\\ Within\end{tabular}}} & \multicolumn{1}{c|}{\textbf{\begin{tabular}[c]{@{}c@{}}Clean\\ Across\end{tabular}}} & \multicolumn{1}{c|}{\textbf{\begin{tabular}[c]{@{}c@{}}Other\\ Within\end{tabular}}} & \textbf{\begin{tabular}[c]{@{}c@{}}Other\\ Across\end{tabular}} \\ \midrule
\textbf{No-samp.} & \multicolumn{1}{c|}{3.6} & \multicolumn{1}{c|}{4.82} & \multicolumn{1}{c|}{5.71} & 9.15 \\
\textbf{No-samp. +  norm.} & \multicolumn{1}{c|}{3.45} & \multicolumn{1}{c|}{4.49} & \multicolumn{1}{c|}{5.48} & 8.74 \\  \midrule
\textbf{Rand.-samp. +  norm.} & \multicolumn{1}{c|}{3.62} & \multicolumn{1}{c|}{4.68} & \multicolumn{1}{c|}{5.71} & 8.63 \\
\textbf{Poisson-samp \cite{sarndal2003model} +  norm.} & \multicolumn{1}{c|}{3.77} & \multicolumn{1}{c|}{5.08} & \multicolumn{1}{c|}{5.56} & 9.01 \\ \midrule
\textbf{N = 10} & \multicolumn{1}{c|}{3.35} & \multicolumn{1}{c|}{4.28} & \multicolumn{1}{c|}{5.25} & 8.36 \\
\textbf{N = 30} & \multicolumn{1}{c|}{\textbf{3.26}} & \multicolumn{1}{c|}{\textbf{4.13}} & \multicolumn{1}{c|}{\textbf{5.00}} & \textbf{8.05} \\
\textbf{N = 50} & \multicolumn{1}{c|}{3.73} & \multicolumn{1}{c|}{4.69} & \multicolumn{1}{c|}{5.62} & 8.88 \\
\textbf{N = 30$^*$} & \multicolumn{1}{c|}{3.93} & \multicolumn{1}{c|}{5.45} & \multicolumn{1}{c|}{5.87} & 9.42 \\ \bottomrule
\end{tabular}
\label{tab:ABX3}
\end{table}
\begin{table}[t!]
\centering
\caption{ABX error (\%) on the ZeroSpeech development dataset {for different number of LSTM layers (L),  hidden units
(H)} and for retaining the top $D$ dimensions (ranked using the boosted decision tree algorithm) of the HUC representations.    { The hyper-parameters, $\lambda$, $k$ and $N$ are set to $1\mathrm{e}{-4}$, $200$ and $30$ respectively. The models in first set of rows were trained on Librispeech $100$h and second set of rows were trained on Librispeech $960$h respectively. ``-+'' indicates that the system configuration  is same as the one in the previous row except for additional configuration indicated after ``+"}}
\begin{tabular}{@{}l|cccc@{}}
\toprule
\multirow{2}{*}{\textbf{{\# Layers (L), Units (H),} }} & \multicolumn{4}{c}{\textbf{ABX}} \\ \cmidrule(l){2-5}
\textbf{\& {Dimensions (D)}} & \textbf{\begin{tabular}[c]{@{}c@{}}Clean\\ Within\end{tabular}} & \textbf{\begin{tabular}[c]{@{}c@{}}Clean\\ Across\end{tabular}} & \textbf{\begin{tabular}[c]{@{}c@{}}Other\\ Within\end{tabular}} & \textbf{\begin{tabular}[c]{@{}c@{}}Other\\ Across\end{tabular}} \\ \midrule
{L = 4, H = 512, D = 512}                                   & \multicolumn{1}{c|}{\textbf{{3.21}}}                                      & \multicolumn{1}{c|}{{4.15}}                                                            & \multicolumn{1}{c|}{{5.03}}    & {\textbf{8.00}}      \\
{L = 2, H = 256, D = 256}                                   & \multicolumn{1}{c|}{{3.26}}& \multicolumn{1}{c|}{\textbf{{4.13}}}                                                  & \multicolumn{1}{c|}{\textbf{{5.00}}}                                                  &{8.05}\\
{-- + masking}  & \multicolumn{1}{c|}{ {4.03}} & \multicolumn{1}{c|}{ {5.81}} & \multicolumn{1}{c|}{ {6.73}} &  {9.81} \\  \midrule
\textbf{L = 2, H = 256, D = 256} & \multicolumn{1}{c|}{3.03} & \multicolumn{1}{c|}{3.70} & \multicolumn{1}{c|}{4.61} & 7.20 \\
\textbf{L = 2, H = 256, D = 224} & \multicolumn{1}{c|}{3.01} & \multicolumn{1}{c|}{3.70} & \multicolumn{1}{c|}{4.59} & 7.04 \\
\textbf{L = 2, H = 256, D = 196} & \multicolumn{1}{c|}{2.92} & \multicolumn{1}{c|}{\textbf{3.50}} & \multicolumn{1}{c|}{\textbf{4.47}} & \textbf{6.95} \\
\textbf{L = 2, H = 256, D = 160} & \multicolumn{1}{c|}{2.96} & \multicolumn{1}{c|}{3.69} & \multicolumn{1}{c|}{4.62} & 7.09 \\
\textbf{L = 2, H = 256, D = 128} & \multicolumn{1}{c|}{\textbf{2.91}} & \multicolumn{1}{c|}{3.65} & \multicolumn{1}{c|}{4.70} & 7.13 \\
\textbf{L = 2, H = 256, D = 96} & \multicolumn{1}{c|}{2.98} & \multicolumn{1}{c|}{3.72} & \multicolumn{1}{c|}{4.80} & 7.43 \\
\textbf{L = 2, H = 256, D = 64} & \multicolumn{1}{c|}{3.43} & \multicolumn{1}{c|}{4.19} & \multicolumn{1}{c|}{5.18} & 8.15 \\
\bottomrule
\end{tabular}
\label{tab:ABX4}
\end{table}

\begin{table*}[t!]
\centering
\caption{ABX result for various systems on the ZeroSpeech $2021$ challenge dataset. The systems mentioned in the first three sets of rows are relatively low GPU budget models ($<200$h). The systems in the second set of rows use the speaker information, while the rest of the models do not utilize this information. The proposed HUC approaches, reported in third set of rows, is also of low GPU budget $<200$h. The systems in the last set of rows use a high GPU budget model ($>1000$h). }
\begin{tabular}{@{}c|cccc||cccc@{}}
\toprule
\multirow{2}{*}{\textbf{System}} & \multicolumn{4}{c|}{\textbf{ABX (dev set)}}                                                                                                                                                                                                                                                                                           & \multicolumn{4}{c}{\textbf{ABX (test set)}}                                                                                                                                                                                                                                                                                           \\ \cmidrule(l){2-9}
                                 & \multicolumn{1}{c|}{\textbf{\begin{tabular}[c]{@{}c@{}}Clean\\ Within\end{tabular}}} & \multicolumn{1}{c|}{\textbf{\begin{tabular}[c]{@{}c@{}}Clean\\ Across\end{tabular}}} & \multicolumn{1}{c|}{\textbf{\begin{tabular}[c]{@{}c@{}}Other\\ Within\end{tabular}}} & \textbf{\begin{tabular}[c]{@{}c@{}}Other\\ Across\end{tabular}} & \multicolumn{1}{c|}{\textbf{\begin{tabular}[c]{@{}c@{}}Clean\\ Within\end{tabular}}} & \multicolumn{1}{c|}{\textbf{\begin{tabular}[c]{@{}c@{}}Clean\\ Across\end{tabular}}} & \multicolumn{1}{c|}{\textbf{\begin{tabular}[c]{@{}c@{}}Other\\ Within\end{tabular}}} & \textbf{\begin{tabular}[c]{@{}c@{}}Other\\ Across\end{tabular}} \\ \midrule
\textbf{CPC-Big Baseline~\cite{dunbar2021zero}}        & \multicolumn{1}{c|}{6.38}                                                            & \multicolumn{1}{c|}{8.26}                                                            & \multicolumn{1}{c|}{10.22}                                                           & 14.86                                                           & \multicolumn{1}{c|}{6.71}                                                            & \multicolumn{1}{c|}{8.41}                                                            & \multicolumn{1}{c|}{10.62}                                                           & 15.06                                                           \\
\textbf{Wav2Vec~\cite{schneider19_interspeech}}                 & \multicolumn{1}{c|}{9.47}                                                            & \multicolumn{1}{c|}{11.69}                                                           & \multicolumn{1}{c|}{12.35}                                                           & 17.61                                                           & \multicolumn{1}{c|}{9.02}                                                            & \multicolumn{1}{c|}{11.33}                                                           & \multicolumn{1}{c|}{12.34}                                                           & 18.44                                                           \\
\textbf{vq-Wav2Vec Gumbel~\cite{baevski2020vqwav2vec}}       & \multicolumn{1}{c|}{10.66}                                                           & \multicolumn{1}{c|}{12.02}                                                           & \multicolumn{1}{c|}{13.17}                                                           & 17.55                                                           & \multicolumn{1}{c|}{10.01}                                                           & \multicolumn{1}{c|}{12.07}                                                           & \multicolumn{1}{c|}{13.39}                                                           & 18.68                                                           \\
\textbf{Maekaku et al.~\cite{maekaku21_interspeech}}         & \multicolumn{1}{c|}{3.28}                                                            & \multicolumn{1}{c|}{4.14}                                                            & \multicolumn{1}{c|}{4.97}                                                            & 8.27                                                            & \multicolumn{1}{c|}{3.15}                                                            & \multicolumn{1}{c|}{4.25}                                                            & \multicolumn{1}{c|}{5.13}                                                            & 8.64                                                            \\
\textbf{Krishna et al.~\cite{varunicassp2022}}          & \multicolumn{1}{c|}{5.83}                                                            & \multicolumn{1}{c|}{8.21}                                                            & \multicolumn{1}{c|}{7.71}                                                            & 13.60                                                           & \multicolumn{1}{c|}{5.04}                                                            & \multicolumn{1}{c|}{7.08}                                                            & \multicolumn{1}{c|}{7.89}                                                            & 14.01                                                           \\
\textbf{Van Nierkerk et al.~\cite{niekerk21_interspeech}}     & \multicolumn{1}{c|}{5.37}                                                            & \multicolumn{1}{c|}{6.63}                                                            & \multicolumn{1}{c|}{8.80}                                                            & 12.89                                                           & \multicolumn{1}{c|}{5.41}                                                            & \multicolumn{1}{c|}{6.89}                                                            & \multicolumn{1}{c|}{8.67}                                                            & 13.14                                                           \\
\midrule
\textbf{Chorowski et al.~\cite{chorowski21_interspeech}}       & \multicolumn{1}{c|}{2.95}                                                            & \multicolumn{1}{c|}{3.60}                                                            & \multicolumn{1}{c|}{4.50}                                                            & 6.99                                                            & \multicolumn{1}{c|}{\textbf{2.85}}                                                            & \multicolumn{1}{c|}{3.69}                                                            & \multicolumn{1}{c|}{4.44}                                                            & 7.28                                                            \\
 \midrule
\textbf{HUC}                      & \multicolumn{1}{c|}{3.26}                                                            & \multicolumn{1}{c|}{4.14}                                                            & \multicolumn{1}{c|}{4.94}                                                            & 7.96                                                            & \multicolumn{1}{c|}{3.25}                                                            & \multicolumn{1}{c|}{4.10}                                                            & \multicolumn{1}{c|}{4.92}                                                            & 8.15                                                            \\
\textbf{ HUC + mean norm.}                 & \multicolumn{1}{c|}{3.17}                                                            & \multicolumn{1}{c|}{3.96}                                                            & \multicolumn{1}{c|}{4.79}                                                            & 7.78                                                            & \multicolumn{1}{c|}{3.09}                                                            & \multicolumn{1}{c|}{3.93}                                                            & \multicolumn{1}{c|}{4.66}                                                            & 8.06                                                            \\
\textbf{-- + data samp.}              & \multicolumn{1}{c|}{3.03}                                                            & \multicolumn{1}{c|}{3.70}                                                            & \multicolumn{1}{c|}{4.61}                                                            & 7.20                                                            & \multicolumn{1}{c|}{2.92}                                                   & \multicolumn{1}{c|}{3.82}                                                            & \multicolumn{1}{c|}{4.59}                                                            & 7.46                                                            \\
\textbf{-- + dim. red.}    & \multicolumn{1}{c|}{\textbf{2.92}}                                                   & \multicolumn{1}{c|}{\textbf{3.50}}                                                   & \multicolumn{1}{c|}{\textbf{4.47}}                                                   & \textbf{6.95}                                                   & \multicolumn{1}{c|}{2.92}                                                   & \multicolumn{1}{c|}{\textbf{3.67}}                                                   & \multicolumn{1}{c|}{\textbf{4.36}}                                                   & {7.17}  \\
\midrule
\midrule
\textbf{wav2Vec 2.0~\cite{baevski2020wav2vec}}             & \multicolumn{1}{c|}{8.48}                                                            & \multicolumn{1}{c|}{9.76}                                                            & \multicolumn{1}{c|}{10.34}                                                           & 14.22                                                           & \multicolumn{1}{c|}{7.75}                                                            & \multicolumn{1}{c|}{9.56}                                                            & \multicolumn{1}{c|}{10.78}                                                           & 15.04                                                           \\
\textbf{{HuBERT~\cite{hsu2021hubert}}}             & \multicolumn{1}{c|}{{3.40}}                                                           & \multicolumn{1}{c|}{{4.16}}                                                            & \multicolumn{1}{c|}{\textbf{ { 4.47 }  }}                                                           & {6.97 }                                                           & \multicolumn{1}{c|}{  3.31  }                                                            & \multicolumn{1}{c|}{ 4.17}                                                            & \multicolumn{1}{c|}{ 4.54}                                                           & \textbf{ 6.98}                                                           \\
\textbf{{Best-RQ~\cite{chiu2022self} }}             & \multicolumn{1}{c|}{{4.25}}                                                           & \multicolumn{1}{c|}{{5.12}}                                                            & \multicolumn{1}{c|}{{ 4.96 } }                                                           & {7.30 }                                                           & \multicolumn{1}{c|}{  4.10  }                                                            & \multicolumn{1}{c|}{ 5.03}                                                            & \multicolumn{1}{c|}{ 5.01}                                                           &  7.19                                                           \\
\bottomrule
\end{tabular}
\label{tab:ABX5}
\end{table*}
\subsection{Hyper-parameter selection}
The results for the phonetic sub-task  experiments on the development set of ZeroSpeech 2021 data are shown in Table~\ref{tab:ABX1}, \ref{tab:ABX2}, \ref{tab:ABX3},  and \ref{tab:ABX4} for various choices of hyper-parameters $\lambda$, $k$, $N$, {architecture} and $D$ respectively. { Unless mentioned explicitly, the best value of a hyper parameter found in an experiment is used for all the remaining experiments}. { The hyper-parameter selection was done by pre-training the models on Librispeech $100$h split with the ABX metric on the ZeroSpeech development set as the performance measure}. The key observations from these experiments are,

\subsubsection{Regularization Parameter $\lambda$} The regularization parameter $\lambda$ controls the amount of categorical learning induced by the cluster loss and the long-term predictive power induced by the CPC loss. As shown in the experiments reported in Table~\ref{tab:ABX1}, the best ABX error is achieved for a  value of $\lambda = 1e-4$. This implies that a higher weight for the HUC loss is more useful for
sub-word discovery task in the ZeroSpeech data. The experiments also highlight that, using the HUC loss alone ($\lambda = 0$), is only slightly inferior to the best results achieved.

\subsubsection{Number of clusters $k$ }
The experiments with different choices for the number of clusters, $k$, are reported in Table~\ref{tab:ABX2}. These results show that $k=200$ achieves the best compromise between the goals of consistency and conciseness.


\subsubsection{Number of pseudo-speaker clusters used in sampling $N$}\label{subsubsec:data_label}
As mentioned in Sec.~\ref{sec:data-sampling}, the pseudo-speaker clusters are identified by clustering the utterance level means of the context vector representations, $\boldsymbol{\Bar  c}^i$. The number of clusters is obtained by the choosing the knee-point\footnote{We compute knee point using \url{https://kneed.readthedocs.io/en/stable/}}. For the Librispeech $100$h dataset, this value is found to be $M=250$. Further, choosing $N$ extreme cluster centroids for data sampling ensures diversity of the data used in the clustering. The number of cluster centroids chosen ($N$) is varied and the ABX results for these experiments are reported in Table \ref{tab:ABX3}. The first row reports the experiments without any data sampling and without the mean normalization step, while the second row reports the results for the same setting with the mean normalization step.  The mean normalization improves the ABX metric for all the conditions of within and across speaker test settings. Further, the next three rows with the data sampling, using the mean normalized context vector representations, show that sampling only from $N=30$ farthest clusters gives additional performance improvements. In order to further illustrate the benefit of choosing the farthest centroids (diverse pseudo-speakers), we also experiment by sampling from $N=30$ nearest centroids. The ABX results show that the error is significantly higher (higher than the case without any sampling) for this experiment indicating that the choice of data selection using the most diverse pseudo-speakers is important for identifying the acoustic units of speech. { We also compare the proposed sampling approach with random and Poisson sampling \cite{sarndal2003model} of the    context vectors (Table \ref{tab:ABX3}). The results show that the random and Poisson sampling do not improve the performance when compared with the system that does not involve any sampling.}
{
\subsubsection{Number and dimension of LSTM layers }
The experiments comparing the network architecture of CPC-big~\cite{dunbar2021zero} using $4$ layers of LSTM with $512$ units and $2$ layers of LSTM network with $256$ units are reported in Table~\ref{tab:ABX4}. The results suggest that using a bigger architecture does not improve the performance. For all the subsequent experiments, we use the HUC model with $2$ layers of LSTM and with $256$ units. }

{
\subsubsection{HUC loss with input masking}
In prior works \cite{baevski2020wav2vec,hsu2021hubert,chiu2022self},  masking the output of the $1$-D convolution layer or masking the raw audio signal was found to be useful. We experiment with a masking strategy similar to HuBERT \cite{hsu2021hubert}, where $8\%$ of the $1$-D convolution encoder output with span of length $10$ is masked. The results reported in Table~\ref{tab:ABX4} (third row) show a significant degradation in the performance with the choice of masking. This may be  due to the use of the LSTM aggregator network instead of the transformer   aggregator networks used in prior works \cite{baevski2020wav2vec,hsu2021hubert,chiu2022self}.
}

\subsubsection{Dimensionality of the final  representations $D$}
We use a a gradient boosted decision tree model\cite{Chen:2016:XST:2939672.2939785} to choose the feature dimensions that are relevant for preserving the separability of the pseudo-phoneme classes. The number of dimensions is varied and these results are reported in Table~\ref{tab:ABX4}. The first row reports the results without any dimensionality reduction.  The best results are achieved for a choice of $D=196$, which corresponds to retaining $75$\% of the dimensions.
\begin{table*}[t!]
\centering
\caption{Performance in terms of various spoken language modeling metrics. The systems mentioned in the first three sets of rows are low budget language models ($< 200h$). The results for the models in the second set of rows use the speaker information, while the rest of the models do not utilize this information. The models in the last set of rows use the BERT based high GPU budget language model ($>1000$h).  }
\begin{tabular}{@{}c|c|c|cc||c|c|cc@{}}
\toprule
\multirow{2}{*}{\textbf{System}} & \multirow{2}{*}{\textbf{\begin{tabular}[c]{@{}c@{}}sWUGGY\\ (dev)\end{tabular}}} & \multirow{2}{*}{\textbf{\begin{tabular}[c]{@{}c@{}}sBLIMP\\ (dev)\end{tabular}}} & \multicolumn{2}{c|}{\textbf{\begin{tabular}[c]{@{}c@{}}sSIMI\\ (dev)\end{tabular}}} & \multirow{2}{*}{\textbf{\begin{tabular}[c]{@{}c@{}}sWUGGY\\ (test)\end{tabular}}} & \multirow{2}{*}{\textbf{\begin{tabular}[c]{@{}c@{}}sBLIMP\\ (test)\end{tabular}}} & \multicolumn{2}{c}{\textbf{\begin{tabular}[c]{@{}c@{}}sSIMI\\ (test)\end{tabular}}} \\ \cmidrule(lr){4-5} \cmidrule(l){8-9}
                                 &                                                                                  &                                                                                  & \multicolumn{1}{c|}{\textbf{librispeech}}            & \textbf{synthetic}           &                                                                                   &                                                                                   & \multicolumn{1}{c|}{\textbf{librispeech}}            & \textbf{synthetic}           \\ \midrule
\textbf{BERT-Small Baseline~\cite{dunbar2021zero}}     & 65.81                                                                            & 52.91                                                                            & \multicolumn{1}{c|}{3.88}                            & 5.56                & 65.94                                                                             & 53.02                                                                             & \multicolumn{1}{c|}{3.02}                            & 0.06                         \\
\textbf{Krishna et al.~\cite{varunicassp2022}}          & 61.20                                                                            & 53.42                                                                            & \multicolumn{1}{c|}{\textbf{10.25}}                  & 3.97                         & 59.87                                                                             & 52.20                                                                             & \multicolumn{1}{c|}{2.07}                            & \textbf{12.73}               \\
\textbf{Maekaku et al.~\cite{maekaku21_interspeech}}         & 66.01                                                                            & 54.15                                                                            & \multicolumn{1}{c|}{5.45}                            & 0.81                         & 66.36                                                                             & 53.88                                                                             & \multicolumn{1}{c|}{1.47}                            & 7.00                            \\

\textbf{Van Nierkerk et al~\cite{niekerk21_interspeech}}      & 72.34                                                                            & 53.95                                                                            & \multicolumn{1}{c|}{7.69}                            & 4.29                         & 72.86                                                                             & 53.59                                                                             & \multicolumn{1}{c|}{1.14}                            & 9.23                         \\
\textbf{wav2vec 2.0~\cite{baevski2020wav2vec}}      & 63.99                                                                            & 52.78                                                                            & \multicolumn{1}{c|}{6.33}                            & 1.26                         & 64.01                                                                             & 53.21                                                                             & \multicolumn{1}{c|}{1.06}                            & 3.67                         \\
\textbf{{HuBERT~\cite{hsu2021hubert}}}      & 70.11                                                                            & 54.54                                                                            & \multicolumn{1}{c|}{3.31}                            & 2.17                         & 69.63                                                                             & 53.95                                                                             & \multicolumn{1}{c|}{0.96}                            & 2.81                        \\
\textbf{{Best-RQ~\cite{chiu2022self} } }      & 68.75                                                                           & 54.16                                                                            & \multicolumn{1}{c|}{4.06}                            & 0.67                         & 69.40                                                                             & 53.90                                                                             & \multicolumn{1}{c|}{1.40}                            & 3.02                        \\
\midrule
\textbf{Chorowski et al.~\cite{chorowski21_interspeech}}        & 74.04                                                                            & 52.97                                                                            & \multicolumn{1}{c|}{4.60}                            & -7.75                        & 72.47                                                                             & 52.55                                                                             & \multicolumn{1}{c|}{0.85}                            & 5.15                         \\
 \midrule
\textbf{HUC}                      & 68.57                                                                            & 53.51                                                                            & \multicolumn{1}{c|}{4.37}                            & 0.44                         & 69.01                                                                             & 53.04                                                                             & \multicolumn{1}{c|}{3.37}                   & 4.74                         \\
\textbf{HUC + mean norm.}                 & 74.32                                                                            & 53.74                                                                            & \multicolumn{1}{c|}{5.83}                            & 0.29                         & 74.53                                                                             & 53.07                                                                             & \multicolumn{1}{c|}{0.17}                            & 5.80                         \\
\textbf{-- + data samp.}              & 74.97                                                                   & 55.01                                                                   & \multicolumn{1}{c|}{7.94}                            & 5.47                         & 75.16                                                                    & 54.84                                                                    & \multicolumn{1}{c|}{1.42}                            & 8.22                         \\
\midrule
\midrule
\textbf{CPC-big + BERT-big baseline~\cite{dunbar2021zero}}       & 75.56                                                                            & 56.14                                                                            & \multicolumn{1}{c|}{8.72}                            & 6.25                         & 75.51                                                                             & 56.16                                                                             & \multicolumn{1}{c|}{1.75}                            & 5.17                         \\
\textbf{HUC+mean-norm+data-samp. + BERT-big}       & \textbf{76.61}                                                                            & \textbf{56.27}                                                                            & \multicolumn{1}{c|}{3.68}                            & \textbf{8.08}                         & \textbf{76.80}                                                                             & \textbf{56.95}                                                                             & \multicolumn{1}{c|}{\textbf{4.08}}                            & 7.74                         \\

\bottomrule
\end{tabular}
\label{tab:lm1}
\end{table*}

\subsection{Comparison with other benchmarks on ABX task}
We compare the proposed work with other previously published benchmarks on the ABX task. The results are reported for  the development set and the official evaluation set (test set) in Table~\ref{tab:ABX5}.

The baseline systems compared are the representations derived from the CPC-Big model (used in the pre-training of the HUC representations)~\cite{dunbar2021zero}, wav2vec~\cite{baevski2020wav2vec}, wav2vec-vq~\cite{baevski2020vqwav2vec} and wav2vec 2.0~\cite{baevski2020wav2vec}.  These approaches require lower GPU budget for training ($< 200$h).
Further, recent works published by Maekaku et. al.~\cite{maekaku21_interspeech}, Nierkerk et. al.~\cite{niekerk21_interspeech} are also relatively lower in GPU budget ($< 200$h). However, the approaches reported in  Chorowski et. al.~\cite{chorowski21_interspeech}    use the speaker information in the training. It is note worthy that the proposed HUC approach is low GPU budget ($150$h) and does  not use any  speaker information.

The proposed HUC approach provides consistent improvements for the both the development and evaluation setting. The model is also seen to improve over other high budget approaches, resulting in state-of-art results on most of these sub-tasks. The best results are achieved for the HUC model with mean normalization, data sampling and dimensionality reduction. In comparison with the CPC big baseline \cite{dunbar2017zero} and our prior work \cite{varunicassp2022}, the proposed HUC model improves the ABX error relatively by $54$\% and $45$\%,   respectively.

\begin{table}[t!]
\centering
{
\caption{{GPU Training budget in hours of various SSL models trained on Librispeech $960$h for $400$k update steps}}
\begin{tabular}{c|c}
\hline
\textbf{Model}                                                                & \textbf{Training Budget(h)} \\ \hline
\textbf{wav2vec 2.0 \cite{baevski2020wav2vec}}                                                          & 1344                        \\
\textbf{HuBERT \cite{hsu2021hubert}}                                                               & 1440                        \\
\textbf{Best-RQ \cite{chiu2022self}}                                                              & 1152                        \\
\textbf{CPC-Baseline \cite{dunbar2021zero}}                                                         & 192                         \\
\textbf{HUC+ mean norm + sampling} & \textbf{152}                         \\ \hline
\end{tabular}
\label{tab:GPU_Budget}
}
\end{table}
\subsection{Language modeling tasks}
The architecture of the BERT language model is the same as the BERT-small model described in \cite{dunbar2021zero}. This language model is trained on 960 hours of LibriSpeech data.
 For the language model based evaluation tasks, we cluster the embeddings ($\boldsymbol{\hat c}_t$) from the trained models into discrete tokens using k-means with $k=200$. The BERT language model is trained on these discrete tokens.  The language model is trained using fairseq tools\footnote{\url{https://github.com/pytorch/fairseq}}.
 The results on the language modeling sub-tasks are shown in Table~\ref{tab:lm1}. The HUC based pseudo-phoneme sequences, modeled with the BERT-big architecture,  provides the best performance on most of the language modeling tasks.

{
\subsection{GPU Budget}
We define the GPU budget of a SSL model as the number of GPU hours required to complete $400$k weight update steps on the Librispeech $960$h dataset. For example if a model takes $7$ days to complete $400$k weight update steps  while training on $8$ GPUs, the GPU budget of the model would be $7 \times 8 \times 24 = 1344$h.
We have compared the GPU budget of various models  in   Table~\ref{tab:GPU_Budget}, where we find the HUC framework to be considerably  efficient. }

{
\section{Robustness of Self-Supervised models}\label{sec:Robustness_experiments}
We follow the approach proposed by Gat et al. \cite{gat2023augmentation} to measure the robustness of the models in the presence of noise and semantically invariant  transformations.
Given a speech signal $x \in \mathbb{R^T}$, we apply basic transformation $g : R^T \mapsto R^T$, such as pitch, reverberation or additive noise to obtain $x^\prime$. Both $x$ and $x^\prime$ are then fed to the SSL model $f : R^T \mapsto R^{T^{\prime}}$. The resulting continuous representations are then fed to the k-means quantizer $E : R^{T^{\prime}} \mapsto \{1....K\}^{T^{\prime}}$  and the discretized sequence is de-duplicated (for example a sequence $12,12,34,34,52$ is converted to $12,34,52$). Here $K$ denotes the number of discrete units. The SSL model $F$ and the quantizer $E$ are trained on the ``clean'' data.
The modified Levenshtein distance \cite{levenshtein1966binary}  ${UED}_{\mathbb{D}}$ between $x$ and $x^\prime$ is the given by,
\begin{equation}\label{eq:UED}
 {UED}_{\mathbb{D}} = \sum_{\x \in \mathbb{D}}\frac{1}{T^{\prime}}LEV((E\circ f)(x),(E\circ f\circ g)(x))
\end{equation}
Here,  $\mathbb{D}$ is   evaluation data and LEV is Levenshtein distance \cite{levenshtein1966binary}. All the SSL models  are pre-trained using Librispeech $960$h dataset, while the k-means is trained on the representations using the   Librispeech $100$h subset. The transformations applied are pitch perturbation, uniformly sampled between scales of $-300$ to $300$, reverberation by uniformly sampling room responses with scale between $0$ to $100$, and additive noise,  sampled   from MUSAN dataset \cite{snyder2015musan} with SNR     between $5$ to $15$ dB. The transformations are implemented using WavAugment\footnote{\url{https://github.com/facebookresearch/WavAugment}}.
The results averaged over $5$ runs are reported in Table~\ref{tab:robust_table}. The results show that the representations from the proposed HUC model elicit significant robustness to noise and other perturbations.
\begin{table}[t!]
\centering
{
\caption{{Modified Levenshtein distance measured on
Librispeech test-split averaged over 5 runs (Lower is better).}}
\begin{tabular}{c|ccc}
\hline
\multirow{2}{*}{\textbf{Model}}                                               & \multicolumn{3}{c}{\textbf{Transformation}}                                                    \\ \cline{2-4}
                                                                              & \multicolumn{1}{c|}{\textbf{Pitch}}  & \multicolumn{1}{c|}{\textbf{Noise}}  & \textbf{Reverb.} \\ \hline
\textbf{CPC Baseline \cite{dunbar2021zero}}                                                         & \multicolumn{1}{c|}{253.00$\pm$0.42 }          & \multicolumn{1}{c|}{185.60$\pm$0.87}          & 186.90$\pm$1.18           \\
\textbf{\begin{tabular}[c]{@{}c@{}}HUC + mean norm\\ + sampling\end{tabular}} & \multicolumn{1}{c|}{\textbf{182.16$\pm$0.45}} & \multicolumn{1}{c|}{\textbf{145.81$\pm$0.66}} & \textbf{148.60$\pm$0.39}  \\
\textbf{HuBERT \cite{hsu2021hubert}}                                                               & \multicolumn{1}{c|}{233.10$\pm$0.56}          & \multicolumn{1}{c|}{173.64$\pm$2.60}          & 183.75$\pm$0.27          \\
\textbf{Wav2Vec 2.0 \cite{baevski2020wav2vec}}                                                          & \multicolumn{1}{c|}{378.39$\pm$0.99}          & \multicolumn{1}{c|}{316.84$\pm$1.93}          & 353.90$\pm$0.61          \\
\textbf{Best-RQ \cite{chiu2022self}}                                                              & \multicolumn{1}{c|}{276.41$\pm$0.28 }          & \multicolumn{1}{c|}{211.14$\pm$0.12}          & 223.73$\pm$0.49           \\ \hline
\end{tabular}
\label{tab:robust_table}
}
\vspace{-0.1in}
\end{table}
}
\section{Low resource speech recognition}\label{sec:asr_experiments}
\subsection{TIMIT phoneme recognition}
\subsubsection{Data}
TIMIT~\cite{garofolo1993darpa} corpus consists of $5$ hours of $16$ kHz sampled English read speech. The manually transcribed time-aligned phonetic and word level transcriptions for each utterance are also available.  It contains recordings from $630$ speakers belonging to $8$ different dialects of American English. The training set consists of $3696$ utterances spoken by $462$ speakers. The development set consists of $400$ utterances spoken by $50$ speakers. The test split contains $192$ utterances from $24$ speakers.

\subsubsection{Experimental set up}\label{subsec:E2E_hy}
In the TIMIT dataset, the phoneme recognition task is designed using an encoder decoder attention based neural network. This model is motivated by the hybrid connectionist temporal cost (CTC) attention network proposed for ASR  by Watanabe et. al.~\cite{watanabe2017hybrid}.
The encoder is a $4$-layer recurrent neural network (RNN) with a hidden dimension of $320$ and the dropout is set at $0.2$. The decoder is an attention decoder containing $1$-layer of LSTM units with a hidden dimension of $320$.
The pre-training for all the representation learning based features is performed on the Librispeech $960$ hour dataset.    For the wav2vec  models we have used the model released by fairseq\footnote{\url{https://github.com/facebookresearch/fairseq/blob/main/examples/wav2vec/README.md}}. Further, the representation learning front-ends are fine-tuned   for the task of phoneme recognition.
\subsubsection{Results}
The phoneme recognition results in terms  of phoneme error rate (PER) are reported in Table ~\ref{tab:PER}.
Among the various benchmarks compared, the wav2vec model gives the best PER. The proposed HUC framework improves significantly over the prior works in terms of PER. The model improves the PER relatively by $9.2$\% and $4.6$\% over the baseline front-ends like the CPC-big  and the wav2vec. Further, the results also highlight the incremental improvements in the  PER results with mean normalization and data sampling approaches proposed in this work for the HUC model, which are consistent with the ABX improvements seen on the ZeroSpeech challenge task.
\begin{table}[t!]
\centering
\caption{Phoneme error rate (PER) \% for development  and test dataset of TIMIT corpus for different  front-ends.}
\begin{tabular}{@{}c|c|c@{}}
\toprule
\textbf{Frontend feat.} & \textbf{Dev.} & \textbf{Test} \\ \midrule
\textbf{fbank-pitch} & 18.2 & 21.0 \\
\textbf{CPC-Big~\cite{dunbar2021zero}} & 13.0 & 14.2 \\
\textbf{wav2Vec~\cite{schneider19_interspeech}} & 12.0 & 13.6 \\
\textbf{wav2Vec Gumbel~\cite{baevski2020vqwav2vec}} & 13.8 & 15.7 \\
\textbf{wav2Vec 2.0~\cite{baevski2020wav2vec}} & 15.3 & 17.1 \\
\textbf{HuBERT~\cite{hsu2021hubert}} & 12.1 & 13.7 \\
\textbf{Best-RQ~\cite{chiu2022self}} & 13.4 & 14.6 \\ \midrule
\textbf{HUC} & 12.8 & 14.1 \\
\textbf{HUC + mean norm. } & 12.2 & 13.7 \\
\textbf{-- + data samp.} & \textbf{11.8} & \textbf{13.0} \\ \bottomrule
\end{tabular}\label{tab:PER}
\end{table}

\subsection{ASR experiments}
\subsubsection{Data} The GramVaani Hindi ASR challenge dataset\footnote{\url{https://sites.google.com/view/gramvaaniasrchallenge/dataset}} comprises of $1100$ hours of telephone quality speech data in Hindi language. In the Track-II of the evaluation, the task entails $1000$ hours of unlabelled audio and $100$ hours of  transcribed speech for training the ASR system. The data set also contains subsets of $5$ hours of development and $3$ hours of evaluation, which are  manually transcribed. The audio is recorded over the telephone channel in varying sampling rates of $8$kHz, $16$kHz, $32$kHz, $44$kHz and $48$kHz. In our experiments, we down sample the entire dataset to $8$kHz.

\subsubsection{Experimental setup} The training objective, based on hybrid CTC attention, is similar to the phoneme recognition setup on the TIMIT dataset. The acoustic model encoder contains $12$ layers of conformer architecture~\cite{gulati2020conformer,guo2021recent}. Further, $8$ layers of transformer architecture are used in the decoder. Both the encoder and decoder layers have the hidden dimension set to $2048$ with the number of attention heads set to $8$. The CNN  kernel size in the conformer is set to $15$. The number of byte pair encoding (BPE) is set to $1000$. The model is trained for $20$ epochs with early stopping.
\begin{table}[t!]
\centering
\caption{Word error rate (WER) \% for the development  and test set of GramVaani dataset. The first and second  row shows the result for filter bank features when the ASR system is trained with $100$h of supervised data or with an additional $200$h of augmented unsupervised data with pseudo labels. The rest of the results in the table follow the latter setting.}
\begin{tabular}{@{}c|c|c@{}}
\toprule
\textbf{Models} & \textbf{Dev WER} & \textbf{Test WER} \\ \midrule
\textbf{fbank-pitch (100h.)\cite{tarunIS2022}} & 38.3 & 36.7 \\
\textbf{fbank-pitch (100h. + 200 h.)} & 38.1 & 36.4 \\ \midrule
\textbf{CPC big \cite{dunbar2021zero}}& 37.0 & 35.5 \\
\textbf{wav2vec~\cite{baevski2020wav2vec}} & 35.3 & 33.9 \\
\textbf{wav2vec-vq~\cite{baevski2020vqwav2vec}} & 36.4 & 34.9\\
\textbf{wav2vec 2.0~\cite{baevski2020wav2vec}} & 34.3 & 34.2 \\
\textbf{HuBERT~\cite{hsu2021hubert}} & 32.7 & 32.1 \\
\textbf{Best-RQ~\cite{chiu2022self}} & 32.8 & 31.9 \\
\midrule
\textbf{HUC [k=$50$]} & 35.6 &  34.6\\
\textbf{HUC [k=$200$]} & 35.2 & 33.7 \\
\textbf{HUC [k=$400$]} & 35.3 & 34.2 \\
\textbf{HUC + data samp. [k=$200$]} & 34.8 & 33.9 \\
\textbf{HUC + mean norm. [k=$200$] } & 33.1 &  32.3\\
\textbf{-- + data samp. [k=$200$]} & \textbf{31.8} &  \textbf{31.0}\\
\bottomrule
\end{tabular}\label{tab:WER}
\end{table}
For the semi-supervised training on $1000$ hours of unlabelled data, we pre-train ASR models with $100$ hours of supervised data. Two separate ASR systems are built using hybrid ASR architecture and E2E ASR architecture. Further, the decoded transcripts from both the systems on the $1000$ hours of unlabeled data are compared and a subset of $200$ hours is selected where the agreement between the hybrid ASR and E2E ASR systems are the highest. More details about the pseudo-label generation and modeling are given in \cite{tarunIS2022}. The final ASR system is trained using $100$ hours of fully transcribed data along with $200$ hours of pseudo-labels. All the experiments use RNN language model trained with the text from the $100$ hours of transcripts.

\subsubsection{Results}  The ASR results, in terms of WER, are reported in Table~\ref{tab:WER}. The first row shows the baseline result using only the $100$ hours of supervised audio, while the second row shows the ASR results for the case when $200$ hours of unsupervised audio with the pseudo labels are added to the training of the ASR models.
{The  baseline ASR system trained with $100$ hours was used to generate the pseudo-labels for the remainder of the $1000$ hours of unlabeled speech, and a selection criterion was used to filter   $200$ hours of reliable transcriptions for the self-training.}
The semi-supervised training provides a minor improvement to the ASR results. Hence, this setting is used for the rest of the experiments.

The pre-trained representation learning frameworks (CPC, wav2vec and HUC models) are trained using the $1100$ hours (supervised + unsupervised pool) of Hindi audio data.
 The representation learning models improve significantly over the mel filterbank baseline features. The best baseline SSL model improves relatively over the mel filterbank features by $7.3$\% and $6.8$\% on the development and test sets respectively.

The HUC model experiments show that, with  $k=200$ clusters improves over other choice of $k=50/400$ clusters. Further, the data sampling and the mean normalization techniques subsequently improve the ASR performance. The final HUC model, which combines mean normalization and data sampling, improves significantly over the baseline mel filterbank and wav2vec features. The model achieves relative improvements of $9.9$\% and $8.6$\% over the wav2vec features on the development and test sets respectively.  The Hindi ASR results also highlight the consistent trends seen in the  experiments reported in the previous sections.
{
\subsection{Statistical Significance Test}
To compare the performance of the HUC model relative to other systems in a statistical sense, we use bootstrap estimate of  confidence intervals  \cite{bisani2004bootstrap}.
Table~\ref{tab:stats_table} shows the analysis of the proposed approach against various baseline systems. The probability of improvement (POI)   for proposed system is noticeably high (above $86$\%) for all the system comparisons. }

\begin{table}[t!]
\centering
{
\caption{{Statistical significance of performance improvements observed for the HUC model with mean norm. and sampling w.r.t.  other models. The confidence interval gives the performance bound, while POI indicates the probability of improvement.}}
\begin{tabular}{c|cc|c}
\hline
\multirow{2}{*}{\textbf{Model1 v/s Model2}} & \multicolumn{2}{c|}{\textbf{Confidence Interval}}                    & \multirow{2}{*}{\textbf{POI}} \\ \cline{2-3}
                                   & \multicolumn{1}{c|}{Model1}             & Model2            &                      \\ \midrule
                                   \multicolumn{4}{c}{GramVaani ASR} \\
                                   \midrule
\textbf{HuBERT vs HUC}                      & \multicolumn{1}{c|}{{[}31.2, 32.8{]}}  & {[}30.9, 32{]}   & 87.2                 \\
\textbf{Wav2Vec2 vs HUC}                   & \multicolumn{1}{c|}{{[}32.5, 35.2{]}} & {[}30.9, 32{]}   & 97.8                 \\
\textbf{Best-RQ vs HUC}                     & \multicolumn{1}{c|}{{[}31.3, 33.0{]}}  & {[}30.9, 32{]}   & 86.7                 \\
\textbf{CPC vs HUC}                         & \multicolumn{1}{c|}{{[}33.8, 36.9{]}}  & {[}30.9, 32{]}   & 99.0                 \\ \midrule
     \multicolumn{4}{c}{TIMIT Phoneme Recog.}\\
                                   \midrule
\textbf{HuBERT vs HUC}                      & \multicolumn{1}{c|}{{[}12.6, 14.4{]}}  & {[}12.2, 13.5{]} & 89.8                 \\
\textbf{Wav2Vec2 vs HUC}                    & \multicolumn{1}{c|}{{[}15.7, 17.3{]}}  & {[}12.2, 13.5{]} & 98.7                 \\
\textbf{Best-RQ vs HUC}                     & \multicolumn{1}{c|}{{[}12.4, 14.7{]}}  & {[}12.2, 13.5{]} & 86.1                 \\
\textbf{CPC vs HUC}                         & \multicolumn{1}{c|}{{[}13.3, 15.0{]}}  & {[}12.2, 13.5{]} & 95.3      \\
\bottomrule
\end{tabular}
\label{tab:stats_table}
}
\end{table}
{
\section{Non-semantic Evaluations}\label{sec:noss}
We investigate the effectiveness of HUC for emotion recognition, speaker and language identification tasks. The models   are pre-trained on the Librispeech $960$h \cite{Panayotov2015LibrispeechAA} dataset. For all the tasks, we average the representations at utterance level and the SSL models are not fine-tuned. A linear SVM \cite{cortes1995support} is used on the pooled representations for downstream tasks. The tasks explored here are our implementations of the sub-tasks involved in the non-semantic speech (NOSS) benchmark \cite{shor2020towards}.

\subsection{Experimental setup}
For speaker identification tasks, the  VoxCeleb-1 \cite{Nagrani17} dataset consisting of $1251$ speakers is used. The accuracy score on the test split of the dataset is used as evaluation metric. We use VoxForge  \cite{Voxforge.org} dataset, consisting of $6$ languages (English, Russian, Spanish, German, French and Italian), for language identification tasks. The CREMA-D \cite{cao2014crema} dataset is used for emotion recognition tasks. The $5$ fold cross validation score was used to evaluate all the systems.

\subsection{Results}
The results  are given in Table~\ref{tab:NOSS}. We see that there is significant drop in the performance of our proposed model across all the non-semantic tasks. This could be due to the mean normalization and the data sampling  that renders the representations invariant to speaker, language and emotion. The HUC model, without the mean normalization (second row), shows significant improvements over the proposed model on these tasks, indicating that the mean   contains significant non-semantic information.}
\begin{table}[t!]
\centering
{
\caption{{Performance of various models on non-semantic speech tasks defined as part of the NOSS benchmark \cite{shor2020towards}.}}
\begin{tabular}{@{}c|ccc@{}}
\toprule
\multirow{2}{*}{\textbf{Model}}                                               & \multicolumn{3}{c}{\textbf{Task}}                                                                                                                                                                                                                                         \\ \cmidrule(l){2-4}
                                                                              & \multicolumn{1}{c|}{\textbf{\begin{tabular}[c]{@{}c@{}}Speaker\\ Identification\end{tabular}}} & \multicolumn{1}{c|}{\textbf{\begin{tabular}[c]{@{}c@{}}Language\\ Identification\end{tabular}}} & \textbf{\begin{tabular}[c]{@{}c@{}}Emotion\\ Recognition\end{tabular}} \\ \midrule
\textbf{CPC Baseline\cite{dunbar2021zero}}                                                         & \multicolumn{1}{c|}{48.37}                                                                     & \multicolumn{1}{c|}{94.8}                                                                       & 80.53                                                                  \\
\textbf{HUC}                                                          & \multicolumn{1}{c|}{49.46}                                                                     & \multicolumn{1}{c|}{94.90}                                                                      & 74.83                                                                  \\
\textbf{\begin{tabular}[c]{@{}c@{}}HUC + mean norm\\ + sampling\end{tabular}} & \multicolumn{1}{c|}{20.11}                                                                     & \multicolumn{1}{c|}{82.13}                                                                      & 68.26                                                                  \\
\textbf{wav2vec 2.0 \cite{baevski2020wav2vec}}                                                          & \multicolumn{1}{c|}{79.18}                                                                     & \multicolumn{1}{c|}{96.56}                                                                      & 78.67                                                                  \\
\textbf{HuBERT \cite{hsu2021hubert}}                                                               & \multicolumn{1}{c|}{\textbf{80.96}}                                                            & \multicolumn{1}{c|}{\textbf{99.51}}                                                             & \textbf{81.31}                                                         \\
\textbf{Best-RQ \cite{chiu2022self}}                                                              & \multicolumn{1}{c|}{71.23}                                                                     & \multicolumn{1}{c|}{92.30}                                                                      & 75.94                                                                  \\ \bottomrule
\end{tabular}
\label{tab:NOSS}
}
\end{table}

\section{Discussion}\label{sec:discussion}
The analysis reported in Fig.~\ref{fig:tSNE} showed that the utterance level mean of the embeddings captured speaker level information.
In order to further quantify the speaker invariance and the phoneme discovery properties of the representations, we design a set of experiments on the  Librispeech $100$h dataset. This dataset consists of $251$ speakers.
The phonetic alignments for these experiments come from pre-trained Kaldi ASR model which is operated in a forced alignment mode to generate phoneme-level transcripts at every frame (sampled every $0.01$s) for the $100$ hours of data.
{The pre-trained CPC model used for the experiments employs within utterance (within speaker) negative sampling scheme.}

 \begin{table}[t!]
\caption{The first two columns indicate the frame-level phoneme classification (higher the better) and speaker classification accuracy (lower the better). The third column represents the cluster purity of the utterance-level means of the context-vectors (lower the better).}
\begin{tabular}{@{}c|c|c|c@{}}
\toprule
\multirow{3}{*}{\textbf{Model}} & \textbf{Phoneme} & \textbf{Speaker} & \textbf{Speaker} \\
& \textbf{Class. Acc.} (\%) & \textbf{Class. Acc.} (\%) & \textbf{Clus. Purity}
\\
& (Frame) & (Frame) & (Utterance) \\

\midrule
\textbf{fbank-pitch} & 40.84 & 22.30  & 0.51 \\
\textbf{CPC-big} & 73.86 & 95.74 & 0.91 \\ \midrule
\textbf{HUC mean-norm.} & 74.64 & 18.06 & 0.34 \\
\textbf{-- + data samp.} & \textbf{75.25} & \textbf{13.4} & \textbf{0.31} \\ \bottomrule
\end{tabular}
\label{tab:analysis}
\end{table}


We train a simple linear classifier for classifying the speakers/phonemes at frame-level with the  acoustic representations (fbank/CPC/HUC). Further, the mean of the utterance level representations are clustered into $251$ clusters. The speaker cluster purity is measured as the ratio of the number of samples in a given cluster  belonging to the most dominant speaker to the total number of samples in the cluster. The phoneme/speaker classification accuracy as well as the speaker cluster purity are reported in Table~\ref{tab:analysis}.

As seen in the Table, the CPC representations have an improved phoneme and speaker classification accuracy compared to the fbank-pitch features with the simple linear classifier. The speaker cluster purity is also high, indicating that the CPC representations encode both the phoneme and  speaker information. This is not desired as the phonetic and spoken language modeling tasks require speaker invariant speech representations.

The HUC model, with mean normalization, improves the phoneme accuracy over the CPC baseline features, while also reducing the speaker classification accuracy. The speaker cluster purity is  seen to be lower for the HUC representations. With the data sampling procedure in the HUC framework, the phoneme classification is improved while the speaker classification is further degraded. The speaker cluster purity measure also reduces, highlighting that the representations from the proposed HUC model capture phoneme level units that are speaker invariant.

{The key benefits and limitations of the proposed HUC framework are summarized below,} \\
\textbf{{Benefits}}:
\begin{enumerate}
    \item  {The ability to  learn semantically rich representations of speech in a self-supervised setting.}
    \item {Enable the formation of robust representations that are less susceptible to speech transformations.}
    \item {Having a lower computational budget compared to other popular SSL approaches.}
    \item {Significantly improved performance on a battery of downstream semantic tasks on ZeroSpeech, phoneme recognition and ASR settings.}
  \end{enumerate}
\textbf{{Limitations}}:
\begin{enumerate}
   \item { The major limitation of the model is that the representations, while encoding the semantic aspects, compromise on encoding the non-semantic aspects of the speech  signal. }
   \item {The other limitation includes the need for stage-wise pre-training of the model. These steps include mean normalization, data sampling for k-means and HUC training. }
\end{enumerate}

\section{Summary}\label{sec:conclusion}
The paper presents a hidden unit clustering approach for representation learning of speech from raw audio data. The model  forces the representations to be more categorical in a pseudo-phoneme space. We propose techniques for improving the speaker invariance, consistency and conciseness of the model representations using mean normalization and data sampling.
The quality of the representations are evaluated using a series of phonetic and linguistic evaluations on the ZeroSpeech challenge sub-tasks. In these experiments, we also establish new state-of-art results with relatively lower computational budget.   The model architecture with the same hyper-parameters are also used in ASR experiments on the TIMIT and GramVaani  datasets. The low resource ASR experiments further highlight the benefits of the proposed representation learning framework. The ASR results illustrate improvements over benchmarks on self-supervised representation learning.

\bibliographystyle{IEEEbib}
\bibliography{main}

\end{document}